\documentclass[aps,twocolumn]{revtex4}%
\usepackage{amsfonts}
\usepackage{amsmath}
\usepackage{amssymb}
\usepackage{graphicx}%
\begin{document}
\title{Effect of weak impurities on electronic properties of graphene: functional
renormalization-group analysis}
\author{A. Katanin}
\affiliation{Institute of Metal Physics, 620990, Ekaterinburg, Russia, and Ural Federal
University, 620002, Ekaterinburg, Russia}

\begin{abstract}
We consider an effect of weak impurities on electronic properties of graphene
within the functional renormalization-group approach. The energy dependences
of the electronic self-energy and density of states near the neutrality point
are discussed. Depending on the symmetry of the impurities, the electronic
damping $\Gamma$ and density of states $\rho$ can deviate substantially from
those given by the self-consistent Born approximation. We investigate the
crossover from the results of the self-consistent Born approximation, which
are valid far from the neutrality point to the strong-coupling (diffusive)
regime near the neutrality point. For impurities, which are diagonal in both,
valley and sublattice indices, we obtain finite density of states at the Fermi
level with the values, which are much bigger than the result of
self-consistent Born approximation.

\end{abstract}
\maketitle

Graphene is a two-dimensional system with the Dirac electronic dispersion,
that possesses unique properties. In particular, the electronic properties of
graphene near the neutrality point remain a challenging problem of
condensed-matter physics. Unlikely many two-dimensional systems, the
conductivity remains finite at the neutrality point\cite{Novoselov}. Although
the density of states (DOS) for systems with Dirac dispersion is expected to
vanish at the neutrality point, finite DOS at the corresponding position of
the Fermi level was observed experimentally in graphene\cite{FiniteDOS}.

The electronic properties of graphene are expected to be strongly influenced
by impurities. Although the chiral disorder does not lead to the
localization\cite{GadeWegner,Gornyi}, this particular type of disorder is
realized only for infinitely strong impurities (e.g. vacancies). The effect of
strong impurities was intensively investigated within both, analytical
\cite{Ostrovsky1} and numerically exact approaches\cite{Katsnelson,Ostrovsky2}.

At the same time, even weak impurities may have non-trivial effect on the
electronic properties of graphene. In particular, the chiral disorder of the
CI symmetry class was argued to yield the energy dependence $\varepsilon
^{1/7}$ of the density of states near the neutrality point\cite{Tsvelik}. This
result is distinctly different from that of the self-consistent Born
approximation (SCBA) \cite{SCBA}, considering only multiple scattering of
electrons on the same impurity, which enlightens the importance of
inter-impurity scattering processes. It was also shown in Ref. \cite{Gornyi}
that for the long-range disorder, which is diagonal in both valley and
sublattice spaces, SCBA predicts much smaller damping of electrons close to
the neutrality point, than expected from other approaches.

The standard renormalization-group (RG) approach of Ref. \cite{Gornyi} allows
to describe inter-impurity scattering, but it treats the ballistic regime of
the flow only. In particular, this approach yields divergence of the vertices
at some critical length scale, which does not allow to describe the crossover
to the strong-coupling (diffusive) regime. Investigation of the diffusive
regime represents, however, an important problem, since it allows to describe
physical properties in the range of fillings close to the neutrality point.
Although some results for the density of states and conductivity were obtained
within the nonlinear-sigma model approach (see, e.g., Refs.
\cite{Tsvelik,AIIIzerobeta,Gornyi} and references therein), it is interesting
to perform the analysis starting from the weak-coupling point of view, which
may allow to treat both the strong-coupling (diffusive) regime of the flow and
the crossover from ballistic to diffusive regime.

In this paper we investigate the effect of long-range and chiral potential
impurities on the electronic properties of graphene within the recently
proposed Wick-ordered functional renormalization group
scheme\cite{Salmhofer,Katanin}, allowing to treat self-energy effects beyond
the leading order of perturbation theory.

We consider quartic interaction between Dirac fermions due to averaged
potential impurity scattering. Assuming only second-order cumulants are
important (i.e the impurity potentials are substantially weak) the
corresponding action reads\cite{Gornyi,Foster}%
\begin{align}
S &  =\int d^{2}x\left[  \int d\tau\overline{\psi}(\gamma_{\mu}\partial_{\mu
})\psi\right.  \label{S2}\\
&  \left.  -\frac{1}{2}n_{\text{imp}}%
{\displaystyle\sum\limits_{l}}
T_{l}^{2}\int d\tau\int d\tau^{\prime}(\overline{\psi}_{\tau}M_{l}\psi_{\tau
})(\overline{\psi}_{\tau^{\prime}}M_{l}\psi_{\tau^{\prime}})\right]  \nonumber
\end{align}
where $\mu=0,1,2,$ $\partial=(\partial_{\tau},v_{F}\partial_{x},v_{F}%
\partial_{y}),$ $\overline{\psi}=\psi^{+}\gamma_{0},\ \gamma_{\mu}$ are the
Dirac matrices, e.g.%
\begin{equation}
\gamma_{0}=\left(
\begin{array}
[c]{cc}%
\sigma_{3} & 0\\
0 & -\sigma_{3}%
\end{array}
\right)  ,\ \ \gamma_{a}=\left(
\begin{array}
[c]{cc}%
\sigma_{a} & 0\\
0 & -\sigma_{a}%
\end{array}
\right)  ,\label{gamma}%
\end{equation}
where $a=1,2,$ $\sigma_{i}$ are the Pauli matrices. The quadratic part of the
action (\ref{S2}) represents the continuum limit of the microscopic
tight-binding model (see, e.g., Refs. \cite{Gornyi,Foster}) and corresponds to
the representation $\psi=\{\psi_{A}^{1},\psi_{B}^{1},\psi_{B}^{2},-\psi
_{A}^{2}\}$ ($\psi_{s}^{m}$ is an annihilation operator for the electron in
valley $m$ and sublattice $s$). The quantities $T_{l}$ represent scattering
amplitudes (impurity potentials) in different channels; the latter are
described by $4\times4$ matrices $M_{l},$ belonging to a linearly independent
set of complex matrices with $(\gamma_{0}M_{l})^{2}=I$ ($I$ - identity
matrix). The interaction in Eq. (\ref{S2}) is obtained after averaging over
impurity positions with $\langle U_{l}(\mathbf{x})U_{l}(\mathbf{x}^{\prime
})\rangle=n_{\text{imp}}T_{l}^{2}\delta^{(2)}(\mathbf{x}-\mathbf{x}^{\prime
}),$ $n_{\text{imp}}$ is the impurity concentration. Eq. (\ref{S2}) neglects
higher than second order scattering processes on the same impurity, which
implies that this equation is applicable in the limit of weak impurities,
$2\pi n_{\text{imp}}[T_{l}/(2\pi v_{F})]^{n}\ll1$ for $n\geq1$.

\textit{The mean-field (self-consistent Born) approximation} for the model
(\ref{S2}) yields for the fermionic self-energy $\widehat{\Sigma}%
^{\varepsilon}$ (see, e.g. Ref. \cite{Gornyi} and references therein)%
\begin{equation}
\widehat{\Sigma}^{\varepsilon}=-n_{\text{imp}}\sum_{\mathbf{k,}l}T_{l}%
^{2}M_{l}\frac{1}{\gamma_{0}\varepsilon-\widehat{\Sigma}^{\varepsilon
}+\mathrm{i}v_{F}\gamma_{a}k_{a}}M_{l}\label{Sigma_TM}%
\end{equation}
Taking into account that $(M_{l}\gamma_{0})^{2}=I$, the solution to this
equation has the form $\widehat{\Sigma}^{\varepsilon}=\gamma_{0}%
\Sigma(\varepsilon)$ with%
\begin{equation}
\Sigma(\varepsilon)=\frac{U^{2}n_{\text{imp}}}{4\pi v_{F}^{2}}[\varepsilon
-\Sigma(\varepsilon)]\ln\frac{-(v_{F}\Lambda_{\mathrm{uv}})^{2}}%
{[\varepsilon-\Sigma(\varepsilon)]^{2}}\label{Sigma_TM0}%
\end{equation}
where $U^{2}=%
{\displaystyle\sum\nolimits_{l}}
T_{l}^{2}$, and $\Lambda_{\mathrm{uv}}$ is an ultraviolet momentum cutoff. At
$\varepsilon\rightarrow0$ equation (\ref{Sigma_TM0}) yields \cite{Gornyi}
\begin{equation}
\Sigma(0)=-\mathrm{i}\Gamma;\ \Gamma\simeq\Lambda_{\mathrm{uv}}e^{-2\pi
v_{F}^{2}/(n_{\text{imp}}U^{2})}\label{Sigma}%
\end{equation}
Note that exponential smallness of $\Gamma$ in graphene in the limit of weak
impurities is due to vanishing of the density of states of pure system at the
Fermi level. The density of states of impure system at the Fermi level in SCBA
reads\cite{Gornyi}
\begin{equation}
\rho(0)=\frac{4\Gamma}{\pi n_{\text{imp}}U^{2}}%
\end{equation}

\textit{The renormalization group. }To treat the effect of weak impurities
beyond SCBA, we apply the recently proposed Wick-ordered functional
renormalization-group scheme\cite{Salmhofer-paper,Salmhofer,Katanin} by
considering the sharp momentum cutoff of the electronic propagator in the
form
\begin{equation}
C_{\Lambda}=\left(  \gamma_{0}\varepsilon-\widehat{\Sigma}_{\Lambda
}^{\varepsilon}+\mathrm{i}v_{F}\gamma_{a}k_{a}\right)  ^{-1}\theta
(|\mathbf{k|}-\Lambda)\label{C_Lambda}%
\end{equation}
where $\Lambda$ is the cutoff parameter. The corresponding single-scale
propagator reads%
\begin{equation}
S_{\Lambda}=-\dot{C}_{\Lambda}+C_{\Lambda}\dot{\widehat{\Sigma}}_{\Lambda
}^{\varepsilon}C_{\Lambda}=\left(  \gamma_{0}\varepsilon-\widehat{\Sigma
}_{\Lambda}^{\varepsilon}+\mathrm{i}v_{F}\gamma_{a}k_{a}\right)  ^{-1}%
\delta(|\mathbf{k}|-\Lambda),\label{S_Lambda}%
\end{equation}
the dot stands for the derivative over $\Lambda$. Following Refs.
\cite{Salmhofer,Katanin}, we choose the Wick-ordering propagator in the form,
which is complementary to $C_{\Lambda}:$%
\begin{equation}
D_{\Lambda}=\left(  \gamma_{0}\varepsilon-\widehat{\Sigma}_{\Lambda
}^{\varepsilon}+\mathrm{i}v_{F}\gamma_{a}k_{a}\right)  ^{-1}\theta
(\Lambda-|\mathbf{k|}).\label{D_Lambda}%
\end{equation}
In the renormalization-group approach we have contribution of three different
channels to the vertex renormalization. The replica trick requires that the
diagrams with closed loops (i.e., those containing summations over number of
fermion species) should be excluded from the diagram technique. The
corresponding RG equations for the vertex $V_{i_{1..4}}^{\varepsilon
\varepsilon^{\prime}}(\mathbf{k}_{1},\mathbf{k}_{2},\mathbf{k}_{3})$ and the
self-energy $\widehat{\Sigma}_{i_{1}i_{3}\Lambda}^{\varepsilon}$
($i_{1,2},\mathbf{k}_{1,2}$ and $i_{3,4},\mathbf{k}_{3,4}$ correspond to
valley-sublattice indices and momenta of incoming and outgoing particles,
$\varepsilon$ and $\varepsilon^{\prime}$ are the energies of the interacting
particles) are presented in the Supplementary material and shown
diagrammatically in Fig. 1.

\begin{figure}[ptb]
\includegraphics[width=8.5cm]{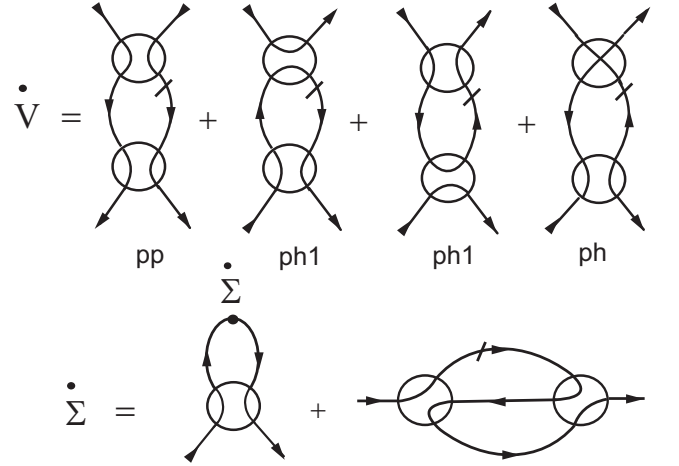}\caption{The diagrammatic form of
the RG equations for the interaction vertices and self-energy. Solid lines
with dash correspond to the single-scale propagator, Eq. (\ref{S_Lambda}),
other solid lines connecting vertices -- to the Wick propagator
(\ref{D_Lambda}). Lines inside vertices connect the legs with the same replica
index. }%
\end{figure}To solve the RG equations numerically we decompose the vertex into
the contribution of particle-hole direct and crossed (ph and ph1), and
particle-particle (pp) channels similarly to Ref. \cite{Salmhofer1},
\begin{align}
V_{i_{1..4}}^{\varepsilon\varepsilon^{\prime}}(\mathbf{k}_{1},\mathbf{k}%
_{2},\mathbf{k}_{3}) &  =V_{i_{1..4}}^{\varepsilon\varepsilon^{\prime
},\mathrm{ph}}(\mathbf{k}_{3}-\mathbf{k}_{2})+V_{i_{1..4}}^{\varepsilon
\varepsilon^{\prime},\mathrm{ph1}}(\mathbf{k}_{1}-\mathbf{k}_{3})\nonumber\\
&  +V_{i_{1..4}}^{\varepsilon\varepsilon^{\prime},\mathrm{pp}}(\mathbf{k}%
_{1}+\mathbf{k}_{2}).\label{Vans}%
\end{align}
Assuming that the disorder is time-reversal invariant, we also use symmetries
of the interaction $V_{i_{1..4}}^{\varepsilon\varepsilon^{\prime},\mathrm{ph}%
}(\mathbf{q})=T_{i_{2}i_{2}^{\prime}}V_{i_{1}i_{2}^{\prime}i_{3}i_{4}^{\prime
}}^{\varepsilon\varepsilon^{\prime},\mathrm{pp}}(\mathbf{q})T_{i_{4}^{\prime
}i_{4}},V_{i_{1..4}}^{\varepsilon\varepsilon^{\prime},\mathrm{ph1}}%
(\mathbf{q})=T_{i_{2}i_{2}^{\prime}}V_{i_{1}i_{2}^{\prime}i_{3}i_{4}^{\prime}%
}^{\varepsilon\varepsilon^{\prime},\mathrm{ph1}}(\mathbf{q})T_{i_{4}^{\prime
}i_{4}}$ where $T=i\gamma_{1}\gamma_{3}$ is the time-inversion matrix,
$\gamma_{3}=\left(
\begin{array}
[c]{cc}%
0 & \mathrm{i}I\\
-\mathrm{i}I & 0
\end{array}
\right)  $ and put $\varepsilon^{\prime}=\varepsilon$, since we consider only
retarted channel of electronic scattering. The initial conditions for the RG
equations read $V_{i_{1..4}}^{\varepsilon\varepsilon}|_{\Lambda=\Lambda
_{\text{uv}}}=n_{\text{imp}}\sum_{l}T_{l}^{2}M_{l}^{i_{1}i_{3}}M_{l}%
^{i_{2}i_{4}};$ $\widehat{\Sigma}_{\Lambda=\Lambda_{\text{uv}}}^{\varepsilon
}=-\mathrm{i}\Gamma\gamma_{0}.$

\textit{Results. }Below we consider the long-range disorder, diagonal in both
sublattice and valley subspaces ($2\times A$II symmetry class), which
corresponds to $M_{1}=\gamma_{0}$, $T_{l}=U\delta_{l,1},$ and one of the
chiral disorders preserving time-reversal symmetry $TM_{l}^{T}T=M_{l}$ with
matrices $M_{l}\in\{\mathrm{i}\gamma_{1,2}\gamma_{3},\mathrm{i}\gamma
_{1,2}\gamma_{5}\}$ describing intervalley scattering of fermions in each of
the sublattices ($\gamma_{5}=-\gamma_{0}\gamma_{1}\gamma_{2}\gamma_{3}$),
$T_{l}=U/2$. Although realistic impurities yield in fact both these types of
disorder (see, e. g., Ref. \cite{Gornyi}), for theoretical analysis it is
informative to consider these types separately. The flow for these types of
the disorder at the scales $\Lambda\gg|\operatorname{Im}\Sigma|/v_{F},$ i.e.
in the ballistic regime, when one can neglect the quasiparticle damping, was
discussed in detail in Ref. \cite{Gornyi} (see also references therein). In
particular, both disorders, $\gamma_{0}$ and i$\gamma_{1,2}\gamma_{3,5}$ yield
the flow of the coupling constants to the strong-coupling regime. For our
numerical calculations we use the parameters $U^{2}n_{\text{imp}}/v_{F}%
^{2}=0.8,$ $\Lambda_{\mathrm{uv}}=2.$ The corresponding SCBA damping is
$\Gamma=7.76\cdot10^{-4}$ $v_{F}$. The results of our fRG analysis are
presented in Figs. 2-5.

\begin{figure}[ptb]
\includegraphics[width=8.5cm]{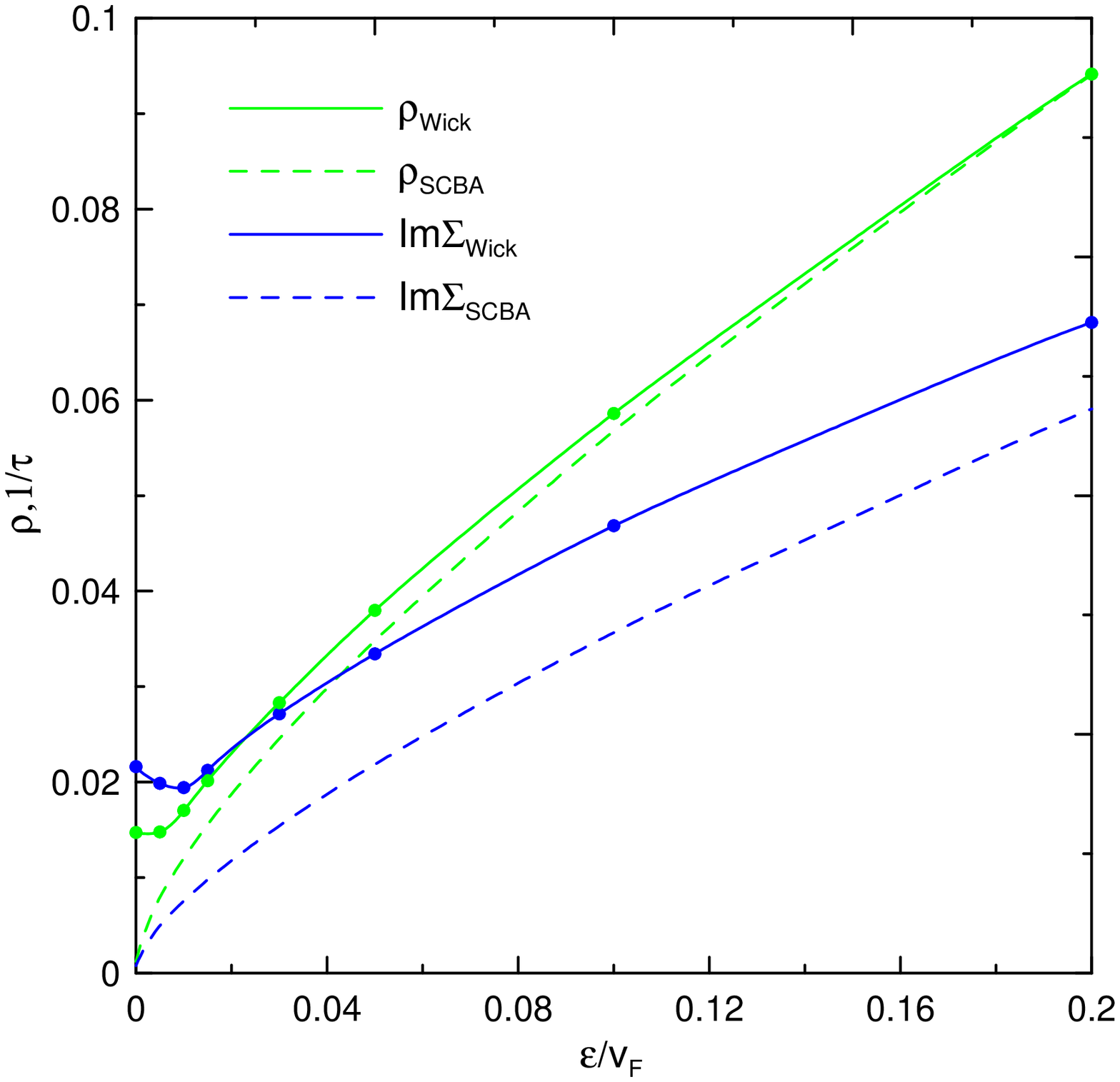}\caption{(Color online) The imaginary
part of the self-energy and the density of states (in units of $v_{F}$) at
different energy distance to the neutrality point for long-range diagonal
disorder. }%
\end{figure}

\begin{figure}[ptb]
\includegraphics[width=8.5cm]{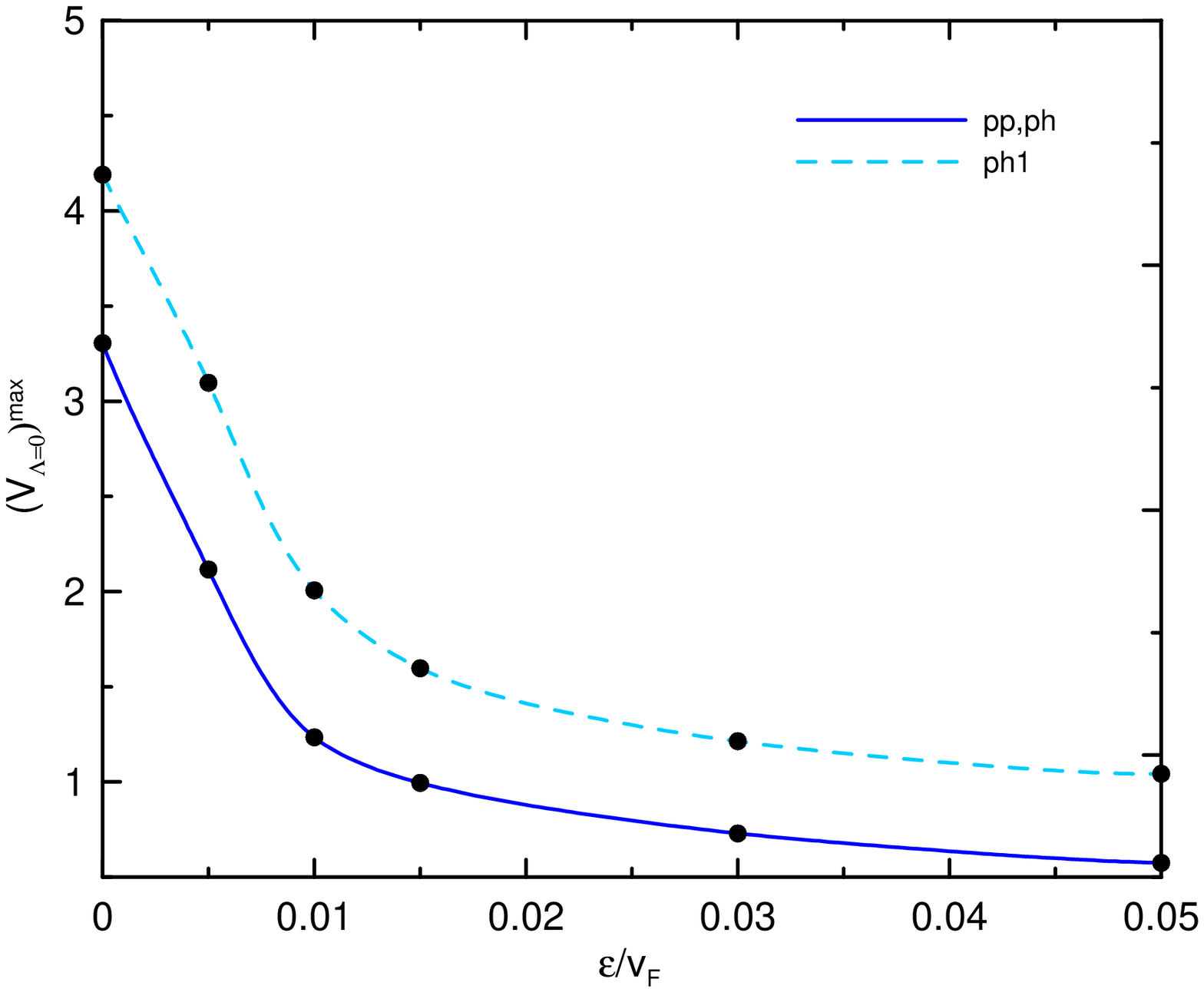}\caption{(Color online) The maximal
interaction vertex in the end of the fRG flow as a function of the energy
distance to the neutrality point for long-range diagonal disorder. }%
\end{figure}\begin{figure}[ptbptb]
\includegraphics[width=8.5cm]{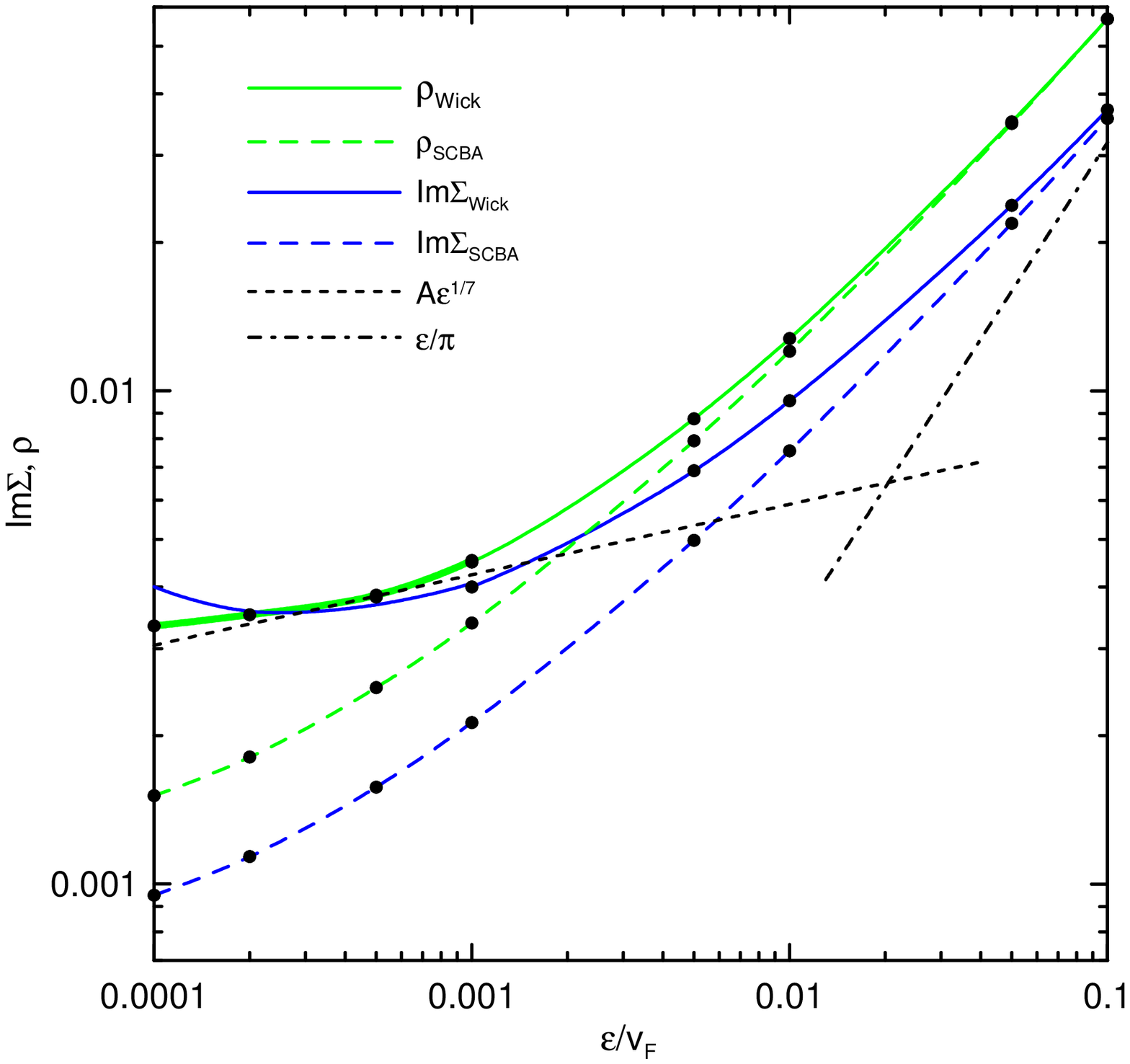}\caption{(Color online) The same as
Fig.1 for chiral disorder of CI symmetry class.}%
\end{figure}

\begin{figure}[ptb]
\includegraphics[width=8.5cm]{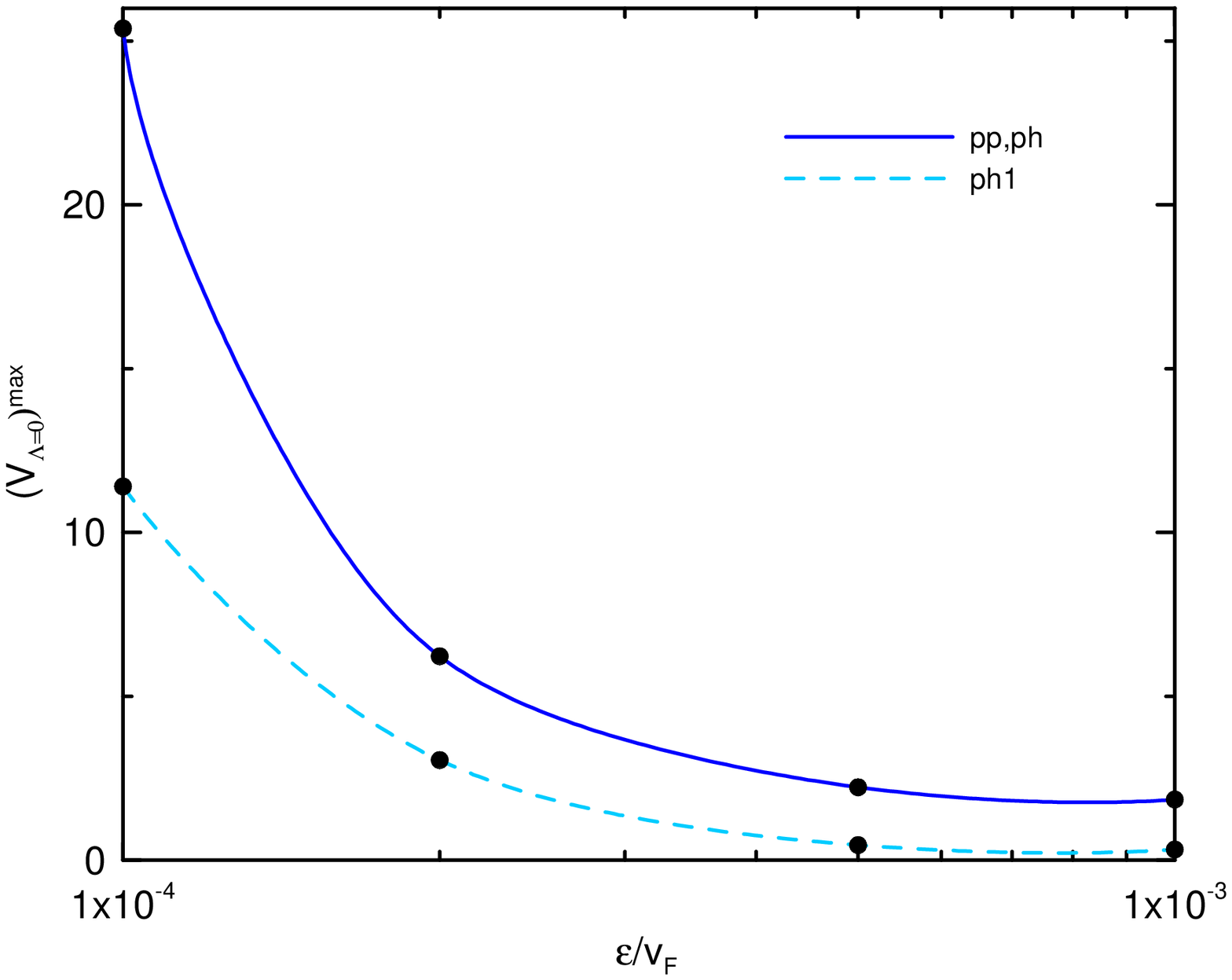}\caption{(Color online) The same as
Fig. 2 for chiral disorder of CI symmetry class}%
\end{figure}

For $\gamma_{0}$ (long-range diagonal) disorder we find saturation of the
density of states and imaginary part of the self-energy at values, which are
much larger than the SCBA results (see Fig. 2). This agrees with earlier
observation in Ref. \cite{Gornyi}, that the critical scale of the divergence
of vertices in the standard RG analysis $\Lambda_{\min}$ differs from the SCBA
result (\ref{Sigma}) by a factor of $2$ in the exponent (for the considered
parameters we find $\Lambda_{\min}\simeq0.04$). Present analysis yields
however finite vertices at and below the critical scale $\Lambda_{\min}$ (Fig.
3). This is due to the cutoff of the divergence of vertices in the present
approach by the quasiparticle damping, which occurs since renormalization of
the vertices involves dressed rather than bare Green functions and reflects
non-perturbative character of the considering approach (cf. the flow into the
magnetic, superconducting, or charge-ordered phases within the one-particle
functional renormalization group approach \cite{Metzner}). From the
field-theoretic point of view, this corresponds to account of many-loop
diagrams within the considered truncation of RG hierarchy. Using SCBA
self-energy as an input of RG flow in the present approach allows us to treat
these self-energy effects more efficiently, since vertices do not grow
strongly already from the beginning of the flow. The finiteness of the
vertices (which divergences are cut at the quasiparticle damping) allows us to
describe the crossover between the ballistic ($|\varepsilon|\gg
|\operatorname{Im}\Sigma|$) and the diffusive ($|\varepsilon|\ll
|\operatorname{Im}\Sigma|$) regimes at the energy $\varepsilon_{\text{cross}%
}\simeq0.02v_{F}\sim\Lambda_{\min}v_{F}$ and obtain density of states and
quasiparticle life-time near the neutrality point. In particular, the
quasiparticle damping approaches the value, which is approximately equal to
the crossover scale $\varepsilon_{\text{cross}}$.

For i$\{\gamma_{1,2}\gamma_{3,5}\}$ disorder we find much smaller values of
the electronic damping, than for the diagonal disorder (Fig. 4). Moreover, at
low energies, logarithm of the density of states scales almost linearly with
$\ln\varepsilon$, with the slope, which is much smaller, than for SCBA result.
The slope of the low-energy region agrees well with that obtained in the
strong-coupling analysis of the problem\cite{Tsvelik}, revealing that the
density of states is expected to behave as $\rho\sim\varepsilon^{1/7}$ at
small energies; for the exponent we obtain in the present approach the value
$0.1$. The values of the maximal vertices, obtained for the considering chiral
disorder, are shown in Fig. 5 and grow on approaching neutrality point. This
behavior of vertices is similar to that found for certain symmetries of the
disorder in $d$-wave superconductors in Ref. \cite{Yash}, and related to
approaching chiral symmetry of the considering model at $\varepsilon=0$, which
implies poles in the the diffusons and cooperons in retarted-retarted (RR)
channel. The obtained momentum dependence of $V^{\varepsilon\varepsilon
,\mathrm{ph}}(\mathbf{k})$ at $\varepsilon>0$ has however a maximum at
$k\sim|\operatorname{Im}\Sigma|/v_{F}$ and approach the corresponding
dependence in the retarted-advanced (RA) channel only for very small
$\varepsilon<10^{-4}v_{F}$. As discussed in Ref. \cite{Yash}, the singularity
of the RR diffusons provides peculiarities of the density of states, which is
likely related to the observed power law behavior of the DOS for CI symmetry. 

\textit{In Conclusion}, we have considered the effect of weak impurities on
electronic properties of graphene. For long-range disorder, we find saturation
of the density of states at the values, which are much bigger than those
obtained previously within the SCBA analysis. On the other hand, for chiral
impurities of CI symmetry class we find indications of vanishing density of
states at the Fermi level, with the power law, which approximates the
previously obtained result $\rho\propto\varepsilon^{1/7}$. The functional
renormalization-group approach allows to describe the crossover from the
ballistic to the diffusive regime in both cases. For realistic impurities,
both long-range and chiral components are present. We expect that long-range
component will be dominating in this case. The qualitative behavior of density
of states in this case agrees with the recent experimental
results\cite{FiniteDOS}.

\textit{Acknowledgements. }The author is grateful to I. V. Gornyi and P.
Ostrovsky for pointing the attention to the problem of weak impurities in
graphene and stimulating discussions on the physics of this material and role
of disorder, and to M. Salmhofer for insightful discussions and hospitality at
the Institute of Theoretical Physics, Heidelberg, Germany.

\newpage

\section*{Supplementary material for the paper "Effect of weak impurities on
electronic properties of graphene: functional renormalization-group analysis"}

In this material we present analytical form of the renormalization-group
equations, shown in Fig. 1 of the paper. The equation for the vertices
$V_{i_{1..4}}^{\varepsilon\varepsilon^{\prime}}(\mathbf{k}_{1},\mathbf{k}%
_{2},\mathbf{k}_{3})$ reads:%

\begin{align}
&
{\displaystyle\sum\limits_{ij}}
\dot{V}_{i_{1..4}}^{\varepsilon\varepsilon^{\prime}}(\mathbf{k}_{1}%
,\mathbf{k}_{2},\mathbf{k}_{3})=\nonumber\\
&  -%
{\displaystyle\sum\limits_{\mathbf{k,}i_{1..4}^{\prime}}}
V_{i_{1}i_{4}^{\prime}i_{1}^{\prime}i_{4}}^{\varepsilon\varepsilon^{\prime}%
}(\mathbf{k}_{1},\mathbf{k+k}_{2}-\mathbf{k}_{3},\mathbf{k})V_{i_{3}^{\prime
}i_{2}i_{3}i_{2}^{\prime}}^{\varepsilon\varepsilon^{\prime}}(\mathbf{k}%
,\mathbf{k}_{2},\mathbf{k}_{3})\nonumber\\
&  \times D_{\mathbf{k}}^{i_{1}^{\prime}i_{3}^{\prime}}(\varepsilon
)S_{\mathbf{k+k}_{\mathbf{3}}-\mathbf{k}_{2}}^{i_{2}^{\prime}i_{4}^{\prime}%
}(\varepsilon^{\prime})-\nonumber\\
&
{\displaystyle\sum\limits_{\mathbf{k,}i_{1..4}^{\prime}}}
\left[  V_{i_{1}i_{1}^{\prime}i_{3}^{\prime}i_{3}}^{\varepsilon\varepsilon
}(\mathbf{k}_{1},\mathbf{k}+\mathbf{k}_{3}-\mathbf{k}_{1},\mathbf{k}%
)V_{i_{4}^{\prime}i_{2}i_{2}^{\prime}i_{4}}^{\varepsilon\varepsilon^{\prime}%
}(\mathbf{k},\mathbf{k}_{2},\mathbf{k+k}_{3}-\mathbf{k}_{1})\right.
\nonumber\\
&  \left.  +V_{i_{1}i_{1}^{\prime}i_{3}i_{3}^{\prime}}^{\varepsilon
\varepsilon^{\prime}}(\mathbf{k}_{1},\mathbf{k}+\mathbf{k}_{3}-\mathbf{k}%
_{1},\mathbf{k}_{3})V_{i_{4}^{\prime}i_{2}i_{4}i_{2}^{\prime}}^{\varepsilon
^{\prime}\varepsilon^{\prime}}(\mathbf{k},\mathbf{k}_{2},\mathbf{k}%
_{1}+\mathbf{k}_{2}-\mathbf{k}_{3})\right]  \nonumber\\
&  \times D_{\mathbf{k}}^{i_{3}^{\prime}i_{4}^{\prime}}(\varepsilon
)S_{\mathbf{k+k}_{1}-\mathbf{k}_{3}}^{i_{2}^{\prime}i_{1}^{\prime}%
}(\varepsilon)\nonumber\\
&  -%
{\displaystyle\sum\limits_{\mathbf{k,}i_{1..4}^{\prime}}}
V_{i_{1}i_{2}i_{1}^{\prime}i_{2}^{\prime}}^{\varepsilon\varepsilon^{\prime}%
}(\mathbf{k}_{1},\mathbf{k}_{2},\mathbf{k})V_{i_{3}^{\prime}i_{4}^{\prime
}i_{3}i_{4}}^{\varepsilon\varepsilon^{\prime}}(\mathbf{k},-\mathbf{k}%
+\mathbf{k}_{1}+\mathbf{k}_{2},\mathbf{k}_{3})\nonumber\\
&  \times D_{\mathbf{k}}^{i_{1}^{\prime}i_{3}^{\prime}}(\varepsilon
)S_{-\mathbf{k+k}_{\mathbf{1}}+\mathbf{k}_{2}}^{i_{2}^{\prime}i_{4}^{\prime}%
}(\varepsilon^{\prime})+S%
\begin{array}
[c]{c}%
\rightleftarrows
\end{array}
D.\label{RGEq}%
\end{align}
where the propagators $S$ and $D$ are defined by the Eqs. (\ref{S_Lambda}) and
(\ref{D_Lambda}) of the main text. For the self-energy correction we obtain
the equation%
\begin{align}
\dot{\widehat{\Sigma}}_{i_{1}i_{3}\Lambda}^{\varepsilon}(\mathbf{k}) &  =%
{\displaystyle\sum\limits_{\mathbf{p,}i_{1..4}^{\prime}}}
V_{i_{1}i_{2}^{\prime}i_{1}^{\prime}i_{3}}^{\varepsilon\varepsilon}%
(\mathbf{k},\mathbf{p},\mathbf{k})D_{\Lambda}^{i_{1}^{\prime}i_{3}^{\prime}%
}(\mathbf{p})\overset{.}{\widehat{\Sigma}}_{i_{3}^{\prime}i_{4}^{\prime
}\Lambda}^{\varepsilon}(\mathbf{p})D_{\Lambda}^{i_{4}^{\prime}i_{2}^{\prime}%
}(\mathbf{p})\nonumber\\
&  -%
{\displaystyle\sum\limits_{\mathbf{pq,}i_{1..6}^{\prime}}}
V_{i_{1}i_{4}^{\prime}i_{1}^{\prime}i_{5}^{\prime}}^{\varepsilon\varepsilon
}(\mathbf{k},\mathbf{q},\mathbf{p})V_{i_{2}^{\prime}i_{6}^{\prime}%
i_{3}^{\prime}i_{3}}^{\varepsilon\varepsilon}(\mathbf{p},\mathbf{k+q-p}%
,\mathbf{q)}\nonumber\\
&  \times S_{\Lambda}^{i_{1}^{\prime}i_{2}^{\prime}}(\mathbf{p})D_{\Lambda
}^{i_{3}^{\prime}i_{4}^{\prime}}(\mathbf{q})D_{\Lambda}^{i_{5}^{\prime}%
i_{6}^{\prime}}(\mathbf{k+q-p})\nonumber\\
&  +2~\text{perm. }S%
\begin{array}
[c]{c}%
\rightleftarrows
\end{array}
D].\label{EqSE}%
\end{align}
Using the ansatz (\ref{Vans}) we split contribution from the particle-hole and
particle particle channels to the vertex evolution. To obtain contribution of
the direct particle-hole channel, we put in Eq. (\ref{RGEq}) $\mathbf{k}%
_{1}=-\mathbf{k}_{2}=\mathbf{k}_{3}=\mathbf{q}/2,$ and obtain%
\begin{align}
&  {\dot{V}}_{i_{1..4}}^{\varepsilon\varepsilon^{\prime},\mathrm{ph}%
}(\mathbf{q})%
\begin{array}
[c]{c}%
=
\end{array}
\nonumber\\
&
{\displaystyle\sum\limits_{i_{1..4}^{\prime},\mathbf{k}}}
\left[  V_{i_{1}i_{4}^{\prime}i_{1}^{\prime}i_{4}}^{\varepsilon\varepsilon
^{\prime},\mathrm{ph}}(\mathbf{q})+V_{i_{1}i_{4}^{\prime}i_{1}^{\prime}i_{4}%
}^{\varepsilon\varepsilon^{\prime},\mathrm{ph1}}(\frac{\mathbf{q}}%
{2}-\mathbf{k)}+V_{i_{1}i_{4}^{\prime}i_{1}^{\prime}i_{4}}^{\varepsilon
\varepsilon^{\prime},\mathrm{pp}}(\mathbf{k-}\frac{\mathbf{q}}{2})\right]
\nonumber\\
&  \times\left[  V_{i_{3}^{\prime}i_{2}i_{3}i_{2}^{\prime}}^{\varepsilon
\varepsilon^{\prime},\mathrm{ph}}(\mathbf{q})+V_{i_{3}^{\prime}i_{2}i_{3}%
i_{2}^{\prime}}^{\varepsilon\varepsilon^{\prime},\mathrm{ph1}}(\mathbf{k-}%
\frac{\mathbf{q}}{2})+V_{i_{3}^{\prime}i_{2}i_{3}i_{2}^{\prime}}%
^{\varepsilon\varepsilon^{\prime},\mathrm{pp}}(\mathbf{k-}\frac{\mathbf{q}}%
{2})\right]  \nonumber\\
&  \times\{S_{\mathbf{k}}^{i_{1}^{\prime}i_{3}^{\prime}}(\varepsilon
)D_{\mathbf{k-q}}^{i_{2}^{\prime}i_{4}^{\prime}}(\varepsilon)+S%
\begin{array}
[c]{c}%
\rightleftarrows
\end{array}
D\};\label{EqVph}%
\end{align}
Putting in Eq. (\ref{RGEq}) $\mathbf{k}_{1}=-\mathbf{k}_{2}=-\mathbf{k}%
_{3}=\mathbf{q}/2,$ we obtain%
\begin{align}
&  {\dot{V}}_{i_{1..4}}^{\varepsilon\varepsilon^{\prime},\mathrm{ph1}%
}(\mathbf{q})%
\begin{array}
[c]{c}%
=
\end{array}
\nonumber\\
&
{\displaystyle\sum\limits_{i_{1..4}^{\prime},\mathbf{k}}}
\left\{  \left[  V_{i_{1}i_{1}^{\prime}i_{3}^{\prime}i_{3}}^{\varepsilon
\varepsilon^{\prime},\mathrm{ph}}(\mathbf{q})+V_{i_{1}i_{1}^{\prime}%
i_{3}^{\prime}i_{3}}^{\varepsilon\varepsilon^{\prime},\mathrm{ph1}}%
(\frac{\mathbf{q}}{2}-\mathbf{k})+V_{i_{1}i_{1}^{\prime}i_{3}^{\prime}i_{3}%
}^{\varepsilon\varepsilon^{\prime},\mathrm{pp}}(\mathbf{k-}\frac{\mathbf{q}%
}{2})\right]  \right.  \nonumber\\
&  \times\left[  V_{i_{4}^{\prime}i_{2}i_{2}^{\prime}i_{4}}^{\varepsilon
\varepsilon^{\prime},\mathrm{ph}}(\mathbf{k-}\frac{\mathbf{q}}{2}%
)+V_{i_{4}^{\prime}i_{2}i_{2}^{\prime}i_{4}}^{\varepsilon\varepsilon^{\prime
},\mathrm{ph1}}(\mathbf{q})+V_{i_{4}^{\prime}i_{2}i_{2}^{\prime}i_{4}%
}^{\varepsilon\varepsilon^{\prime},\mathrm{pp}}(\mathbf{k-}\frac{\mathbf{q}%
}{2})\right]  \nonumber\\
&  +\left[  V_{i_{1}i_{1}^{\prime}i_{3}i_{3}^{\prime}}^{\varepsilon
\varepsilon^{\prime},\mathrm{ph}}(\frac{\mathbf{q}}{2}-\mathbf{k}%
)+V_{i_{1}i_{1}^{\prime}i_{3}i_{3}^{\prime}}^{\varepsilon\varepsilon^{\prime
},\mathrm{ph1}}(\mathbf{q})+V_{i_{1}i_{1}^{\prime}i_{3}i_{3}^{\prime}%
}^{\varepsilon\varepsilon^{\prime},\mathrm{pp}}(\mathbf{k-}\frac{\mathbf{q}%
}{2})\right]  \nonumber\\
&  \left.  \times\left[  V_{i_{4}^{\prime}i_{2}i_{4}i_{2}^{\prime}%
}^{\varepsilon\varepsilon^{\prime},\mathrm{ph}}(\mathbf{q})+V_{i_{4}^{\prime
}i_{2}i_{4}i_{2}^{\prime}}^{\varepsilon\varepsilon^{\prime},\mathrm{ph1}%
}(\mathbf{k-}\frac{\mathbf{q}}{2})+V_{i_{4}^{\prime}i_{2}i_{4}i_{2}^{\prime}%
}^{\varepsilon\varepsilon^{\prime},\mathrm{pp}}(\mathbf{k-}\frac{\mathbf{q}%
}{2})\right]  \right\}  \nonumber\\
&  \times\{S_{\mathbf{k}}^{i_{3}^{\prime}i_{4}^{\prime}}(\varepsilon
)D_{\mathbf{k-q}}^{i_{2}^{\prime}i_{1}^{\prime}}(\varepsilon)+S%
\begin{array}
[c]{c}%
\rightleftarrows
\end{array}
D\};\label{EqVph1}%
\end{align}
Finally, putting in Eq. (\ref{RGEq}) $\mathbf{k}_{1}=\mathbf{k}_{2}%
=\mathbf{k}_{3}=\mathbf{q}/2,$ we obtain%
\begin{align}
&  {\dot{V}}_{i_{1..4}}^{\varepsilon\varepsilon^{\prime},\mathrm{pp}%
}(\mathbf{q})=\nonumber\\
&
{\displaystyle\sum\limits_{i_{1..4},\mathbf{k}}}
\left[  V_{i_{1}i_{2}i_{1}^{\prime}i_{2}^{\prime}}^{\varepsilon\varepsilon
^{\prime},\mathrm{ph}}(\mathbf{k-}\frac{\mathbf{q}}{2})+V_{i_{1}i_{2}%
i_{1}^{\prime}i_{2}^{\prime}}^{\varepsilon\varepsilon^{\prime},\mathrm{ph1}%
}(\frac{\mathbf{q}}{2}-\mathbf{k})+V_{i_{1}i_{2}i_{1}^{\prime}i_{2}^{\prime}%
}^{\varepsilon\varepsilon^{\prime},\mathrm{pp}}(\mathbf{q})\right]
\nonumber\\
&  \times\left[  V_{i_{3}^{\prime}i_{4}^{\prime}i_{3}i_{4}}^{\varepsilon
\varepsilon^{\prime},\mathrm{ph}}(\mathbf{k-}\frac{\mathbf{q}}{2}%
)+V_{i_{3}^{\prime}i_{4}^{\prime}i_{3}i_{4}}^{\varepsilon\varepsilon^{\prime
},\mathrm{ph1}}(\mathbf{k-}\frac{\mathbf{q}}{2})+V_{i_{3}^{\prime}%
i_{4}^{\prime}i_{3}i_{4}}^{\varepsilon\varepsilon^{\prime},\mathrm{pp}%
}(\mathbf{q})\right]  \nonumber\\
&  \times\{S_{\mathbf{k}}^{i_{1}^{\prime}i_{3}^{\prime}}(\varepsilon
)D_{-\mathbf{k+q}}^{i_{2}^{\prime}i_{4}^{\prime}}(\varepsilon^{\prime})+S%
\begin{array}
[c]{c}%
\rightleftarrows
\end{array}
D\}\label{EqVpp}%
\end{align}
The number of vertices per every set $\varepsilon,\varepsilon^{\prime}$ is
$3\cdot256\cdot n_{k}$ where $n_{k}$ is the number of the considered
$k$-points. The equations (\ref{EqSE})-(\ref{EqVpp}) have to be solved
numerically. For the numerical solution, we expand vertices $V_{i_{1..4}%
}^{\varepsilon\varepsilon^{\prime},c}(\mathbf{q})$ ($c=$ph, ph1, or pp) in
each of the channels in harmonics%

\begin{align}
V_{i_{1..4}}^{\varepsilon\varepsilon^{\prime},c}(\mathbf{q})  &
=\sum\limits_{m=0}^{n}\left[  F_{i_{1..4}}^{\varepsilon\varepsilon^{\prime
},cm}(q)\cos(m\varphi_{\mathbf{q}})\right. \nonumber\\
&  \left.  +G_{i_{1..4}}^{\varepsilon\varepsilon^{\prime},cm}(q)\sin
(m\varphi_{\mathbf{q}})\right]  \label{Vharm}%
\end{align}
where $\varphi_{\mathbf{q}}$ is the polar angle of the vector $\mathbf{q}$. In
the numerical solution we take $12$ points in radial $q$ direction and account
for $n+1=3$ harmonics in Eq. (\ref{Vharm}). The results of the solution of
Eqs. (\ref{EqVph})-(\ref{Vharm}) are discussed in the main text.

\end{document}